\documentclass{aipproc}
\layoutstyle{6x9}

\usepackage{amsmath}
\usepackage{graphics}

\begin{document}

\title{Filter Out High Frequency Noise in EEG
Data Using The Method of Maximum Entropy}



\author{Chih-Yuan Tseng}{
  address={
Department of Physics and $^\dagger$Graduate Institute of Systems Biology and Bioinformatics\\
Computational Biology Laboratory\\
National Central University, Chungli, Taiwan 320, ROC}
}

\author{HC Lee$^\dagger$}{
  address={
Department of Physics and $^\dagger$Graduate Institute of Systems Biology and Bioinformatics\\
Computational Biology Laboratory\\
National Central University, Chungli, Taiwan 320, ROC}
}

\begin{abstract}
We propose a maximum entropy (ME) based approach to smooth noise not only in data but also 
to noise amplified by second order derivative calculation of the data especially for electroencephalography 
(EEG) studies. The approach includes two steps, applying method of ME to generate a family of filters and 
minimizing noise variance after applying these filters on data selects the preferred one within the family. 
We examine performance of the ME filter through frequency and noise variance analysis and compare it with 
other well known filters developed in the EEG studies. The results show the ME filters to outperform others. 
Although we only demonstrate a filter design especially for second order derivative of EEG data, these studies 
still shed an informatic approach of systematically designing a filter for specific purposes.
\end{abstract}

\maketitle

\section{Introduction}

Filtering out high frequency noise becomes extremely crucial 
in source localization problem of invasive EEG studies. 
Based on Freeman and Nicholson's works, source localization requires calculation of 
second order derivative of the EEG data \cite{Freeman75a, Freeman75}.  
Because the derivative amplifies especially high frequency noise, Freeman and Nicholson 
proposed to introduce low-pass non-recursive filters to smooth it. 
\cite{Freeman75}. Suppose a set of discrete 
data, $\Phi \left( r\right) $ labeled
by spatial position r, which are distorted by noises $\epsilon _{r}$ and these data are
equally spaced sampled with interval h. The simplest non-recursive
filter is then defined by simply averaging data at r and its neighboring data
with specific set of weighting values $p_{l}$,
\begin{equation}
\Phi _{s}\left( r\right) =\sum_{l=-z}^{z}p_{l}\Phi \left( r+lh\right) \text{
,}  \label{Phi_s-general}
\end{equation}%
where $z$ is an integer and $2z+1$ data points are considered. We will designate this $2z+1$-point 
spatial filter as a S$_{\left( \text{2z+1}\right)}$ filter.
 The mean filter, in which $p_{l}$ 
equals to 1/(2z+1), is one filter 
widely used in many fields for data smoothing. 
Its applicability, 
however, is sometimes limited to data types \cite{Hamming89}. 

Many methods have been developed to design low-pass filters 
in many fields. For example, there are adaptive beam-forming 
(\cite{Borgiotti79} - \cite{Sekigara01}), wavelet
de-noising \cite{Sardy01}, and entropic method (\cite{Avesta04} - \cite%
{DeBrunner04}) etc.. Yet no explicit discussions are given on smoothing noise in second order derivative of contaminated EEG data since Freeman and Nicholson's pioneering works. 
Freeman and Nicholson's approach will be briefly discussed in 
Sec. 2. Two drawbacks, requirements of 
some empirical guidelines and low accuracy of second 
order derivative, in their approach will also be 
addressed. 

Our goal is to provide a ME based approach to filter out high frequency noise especially in 
second order derivative of contaminated EEG data. This 
approach eliminates the two drawbacks in Freeman and Nicholson's approach and 
provides a systematic and robust method for smoothing noise. 
The ME approach and its application in the EEG studies will be presented in Sec. 3 and 4 
respectively. We thereafter examine the proposed filter's performance and compare 
to some other filters such as the mean filter, Freeman and Nicholson's \cite{Freeman75}, and Rappelsberger's \cite{Rappelsberger81} filters through frequency 
and noise variance analysis in Sec. 5. A conclusion is given at last.

\section{Source localization from the invasive EEG data}

In neurophysiology, electric field potentials are initiated during neural activity when ions flow 
across localized regions of cell membranes. These ionic 
flows establish a distribution of current sources and sinks in the extracellular space 
\cite{Freeman75a, Freeman75, Mitzdorf85}. These field 
potentials can be measured in extracellular space 
discretely by the invasive EEG instrument, which usually 
has 16 to 32 detectors on its probe.

According to theory of electromagnetism, the current sources and sinks are associated with 
second order derivative of the field potentials. 
Freeman and Nicholson proposed the current source density (CSD) method
for source localization \cite{Freeman75a, Freeman75}. The CSD 
computes second order derivative approximately through substituting specific 
smoothed $\Phi _{s}\left( r\right)$, Eq. (\ref{Phi_s-general}), into 
Ta$_{3}(r)=\sigma_{c}/h^{2}\left(\Phi \left( r-h\right)-2 \Phi \left( r\right)+\Phi \left( r+h\right)\right)$, 
where three data are considered, $\sigma _{c}$ is conductivity, and $h$ is 
separating distance between two sensors.
A final expression of an approximated second order derivative is given by 
\begin{equation}
D_{2}\left( r\right) = \sigma _{c}/h^{2}\sum_{m=-M}^{M}a_{m}\Phi \left( r+mh\right) \text{, }
\label{D2-general}
\end{equation}%
from 2M+1 datum with weightings $a_{\pm M}=p_{z}$, 
$a_{\pm M \mp 1}=p_{z-1}-2p_{z}$, and $a_{\pm M \mp k}=p_{z-k}-2p_{z-k+1}+p_{z-k+2}$, 
in which $k \geq 2$. 
The preferred smoothing function should be able to reduce high frequency noise 
and noise amplified by operation of the Ta$_{3}(r)$. Freeman and Nicholson
devised several smoothing functions as shown in Table. \ref{Compare filter}, 
denoted by FNSn (\cite{Freeman75}), where Sn stands for n-point spatial smoothing. 
Notes that the FNS$_{\text{3}}$(2) is still used in 
Karwoski et al. \cite{Karwoski96} and Ulbert et al. \cite{Ulbert01}. 
These $p_{l}$ are derived from the least square method \cite{Lanczos56, Hamming71}. 
In addition, they also introduce some ad hoc rules to obtain 5 and 7-point filters. For example, 
the FNS$_{\text{5}}$ is obtained by smoothing data twice using set of (3/10, 4/10 ,3/10). 
Later, several successors have designed other smoothing functions according to their 
needs such as the Rappelsberger's RS$_{\text{3}}$ filter \cite{Rappelsberger81}.  
 
There are two drawbacks in Freeman and Nicholson's CSD method. The first is requirements 
of some ad hoc empirical guidelines for designing smoothing functions such as FNS$_{\text{3}}$(1) or RS$_{\text{3}}$ shown in Table. \ref{Compare filter}. These empirical guidelines may limit the use of these smoothing functions to some specific EEG data. 
For example, Freeman and Nicholson's smoothing functions are suited for Anuran cerebellum 
studies \cite{Freeman75}. The second is these smoothing
functions are devised for second order derivative calculated via Ta$_{3}(r)$, 
which is not an accurate calculation.  In next section, we propose an informatic 
approach based on the ME method to eliminate these two drawbacks.

\begin{table}[!hbp]
\caption{Filters in EEG studies. The table lists five 
smoothing functions \cite{Freeman75, Rappelsberger81}.  
MF: mean filter, FN: Freeman and Nicholson, R: Rappelsberger. }
\label{Compare filter}
\begin{tabular}{|l|cccc|}
\hline
& $p_{0}$ & $p_{\pm 1}$ & $p_{\pm 2}$ & $p_{\pm 3}$  \\ \hline
MFS$_{\text{3}}$    & 0.33 & 0.33 & 0    & 0      \\ 
FNS$_{\text{3}}$(1) & 0.5  & 0.25 & 0    & 0      \\ 
FNS$_{\text{3}}$(2) & 0.43 & 0.29 & 0    & 0     \\ 
RS$_{\text{3}}$     & 0.54 & 0.23 & 0    & 0      \\ 
FNS5    & 0.34 & 0.24 & 0.09 & 0      \\ 
FNS7    & 0.26 & 0.21 & 0.12 & 0.04  \\ \hline
\end{tabular}%
\end{table}

\section{A maximum entropy low-pass filter}


\noindent \textbf{Rationale.} The method of ME is developed as a tool for assigning 
a probability distribution of observing a system to be at a certain state according to information in hand, which is in the form of constraints \cite{Jaynes57}. Based on conventional 
studies on properties of the noise filter, Eq. (\ref{Phi_s-general}), in the EEG 
source localization problems, which will be illustrated later, one can treat the 
weightings $p_l$ as a probability distribution. Thus the ME provides a 
robust and objective approach for determining a preferred set of $p_l$ once constraints 
relevant to noise reduction in the EEG problems are given. 

\noindent \textbf{The constraints.} Conventional studies in  
EEG data analysis have shown some properties that are relevant to smooth noise
using Eq. (\ref{Phi_s-general}). 
First, the $p_{l}$ are positive decimal values and symmetric around $l=0$. 
These values obey a normalization condition $\sum_{l=-z}^{z}p_{l}=1$. Second, 
the preferred set of the $p_{l}$ should have noise in data being mostly removed through 
the $\Phi _{s}\left( r\right)$. 
We expand right hand side of Eq.(\ref%
{Phi_s-general}) with respect to r according to the Taylor expansion and 
use $\sum_{l=-z}^{z}p_{l}=1 $ and $p_{l}=p_{-l}$, which results in the odd order 
derivative terms in the expansion vanished. The final result is 
$\Phi _{s}\left( r\right) =\Phi \left( r\right) +\Phi ^{\left( 2\right)
}\left( r\right) h^{2}\delta_{2}/2+\cdots $, 
where $\Phi ^{\left( 2\right) }\left( r\right) $ denotes exact second order
derivative of clean data and so on and $\delta _{2}=\sum_{l=-z}^{z}p_{l}l^{2}$.
It indicates that smoothing ability is actually related to the 
$\delta _{2}$ value. Summarizing the above results, 
suggests the weightings $p_{l}$ to be a probability distribution of having the 
Eq.(\ref{Phi_s-general}) to reduce noise by the $\delta _{2}$ approximately. 
Thus the ME can be applied to 
determine the preferred set of $p_{l}$.

\noindent \textbf{The preferred low-pass filter for data smoothing.}
The ME states that maximizing 
entropy,  
$S=-\sum_{l=-z}^{z}p_{l}\ln p_{l}$,   
subject to the two constraints, the normalization condition of $p_{l}$ and $\delta _{2}=\sum_{l=-z}^{z}p_{l}l^{2}$ gives a preferred form of the $p_{l}$, 
\begin{equation}
p_{l}=Z^{-1}\exp (-\alpha l^{2})\text{ },  \label{p-value}
\end{equation}%
where $\alpha $ is a Lagrangian multiplier and partition function $%
Z=\sum_{l=-z}^{z}\exp (-\alpha l^{2})$. The preferred set of $p_{l}$ is 
a function of $\alpha $. Namely, the ME method generates a family of $%
p_{l}\left( \alpha \right) $ distribution. In principle, the Lagrangian multiplier $\alpha $ 
can be determined by substituting Eq. (\ref{p-value}) back into 
$\delta _{2}=\sum_{l=-z}^{z}p_{l}l^{2}$, which gives 
$\delta _{2}=-Z^{-1}\partial Z/\partial \alpha $,
if the $\delta _{2}$ value is given. Unfortunately, 
the $\delta _{2}$ value is unknown at this point. Thus, 
we propose an alternative method to determine the $\alpha $.

Since the preferred $\alpha $ value should result in 
corresponding set of the $p_{l}$ to mostly reduce noise, if one can
quantify noise the preferred $\alpha $ value can be determined. 
Although it is difficult to quantify noise, 
noise variance can be quantified by
the autocorrelation function \cite{Hamming89, Feldman91}. 
Suppose field potentials $\Phi \left( r\right) =\Phi _{0}\left( r\right) +
\epsilon _{r}$ is repeatedly measured several times, in which a Gaussian white noise $\epsilon _{r}$ with variance $\sigma $ is considered. For white noise, we have $\left\langle \epsilon _{r}\right\rangle =0$. 
 Besides, noises are uncorrelated at different data points r 
and r$^{\prime}$, $\left\langle \epsilon _{r}\epsilon _{r^{\prime }}\right\rangle =
 \sigma ^{2}$ for $r=r^{\prime }$ and 0 for $r\neq r^{\prime }$ \cite{Hamming89}.
Because noise has zero mean, mean $\left\langle \Phi _{0}\left( r\right) \right\rangle $
is identical to $\Phi _{0}\left( r\right) $, noise-free data. Thus
one can find autocorrelation of $\Phi \left( r\right)$ is   
$C_{o}=\left\langle \left( \Phi \left( r\right) -\left\langle \Phi \left(
r\right) \right\rangle \right) ^{2}\right\rangle =\left\langle \left(
\epsilon _{r}-\left\langle \epsilon _{r}\right\rangle \right)
^{2}\right\rangle =\sigma ^{2}\text{ }$,
the noise variance. Similarly, one can compute the 
autocorrelation function of $\Phi _{s}\left( r\right) $, 
$C_{\text{smooth}}=\left\langle \left( \Phi _{s}\left( r\right) -\left\langle
\Phi _{s}\left( r\right) \right\rangle \right) ^{2}\right\rangle =\sigma
^{2}\sum_{l=-z}^{z}p_{l}^{2}$.  
It indicates noise variance being amplified by $%
\sum_{l=-z}^{z}p_{l}^{2}$ after data is processed by the Eq.(\ref{Phi_s-general}). 
Thus the preferred $\alpha $ value should minimize the noise variance $C_{\text{smooth}}$, which is found to be zero and 
$p_{l}\left(\alpha =0\right) =1/(2z+1)$. This is exactly the mean filter obtained from the 
least square method \cite{Hamming89, Lanczos56}. Yet the ME approach roots in reducing noise variance $C_{\text{smooth}}$ mostly, which is not clear in the least square method. 

\section{Filter out noise in source localization}
\noindent \textbf{Basic features.} Freeman and 
Nicholson suggest to identify current sources and sinks through 
calculating second order derivative of the field potential via the Ta$_{\text{3}}(r)$ with 
specific smoothed $\Phi_{s}(r)$.  
However, the Ta$_{\text{3}}(r)$ does not provide accurate calculation. 
One straightforward way to raise the accuracy is to incorporate more data points into
for calculation,
\begin{equation}
\text{Ta}_{2M+1}\left( r\right) =\sum_{m=-M}^{M}a^{T}_{m}\Phi \left( r+mh\right) \text{, }
\label{general-Ta_2M+1}
\end{equation}
where 2M+1 data points are considered. To determine coefficients $a^{T}_{m}$, one first 
expand $\Phi \left( r+mh\right)$ with respect to r based on the Taylor expansion,  
$\text{Ta}_{2M+1}\left( r\right) =\sum_{m=-M}^{M}a^{T}_{m}\Phi \left( r\right)
+\sum_{m=-M}^{M}a^{T}_{m}mh\Phi ^{\left( 1\right) }\left( r\right)
+\sum_{m=-M}^{M}a^{T}_{m}\left( mh\right) ^{2}/2\Phi ^{\left( 2\right)
}\left( r\right) +\cdots$.   
By requesting coefficient of second order term, $%
\sum_{m=-M}^{M}a^{T}_{m}\left( mh\right) ^{2}/2=1$, and 
rests are zero, we have approximated Ta$_{2M+1}\left(
r\right)$ to be identical to exact second order derivative 
$\Phi ^{\left( 2\right) }\left( r\right) $. 
One then can solve for coefficients $a^{T}_{m}$ given these criteria. 
For three point calculation, coefficients are given in the Ta$_{3}$. 
Coefficients $a^{\text{T}}_{\pm2}=-1/12$, $a^{\text{T}}_{\pm1}=16/12$, 
and $a^{\text{T}}_{0}=-30/12$ for 
Ta$_{5}$ and $a^{\text{T}}_{\pm3}=1/90$, $a^{\text{T}}_{\pm2}=-13.5/90$, 
$a^{\text{T}}_{\pm1}=135/90$, and $a^{\text{T}}_{0}=-245/90$ for Ta$_{7}$.  


Given these coefficients, one can immediately find the 
noise variance to be amplified by 
$C_{D_{2}}=\left\langle \left( D_{2}\left( r\right) -\left\langle D_{2}\left(
r\right) \right\rangle \right) ^{2}\right\rangle =\sigma
^{2}\sum_{m=-M}^{M}(a^{\text{T}}_{m})^{2}$  
after operation of the Ta$_{2M+1}\left(r\right)$. For example, it is 6, 8.03, 9.75 times of the original noise variance for using Ta$_{3}\left(r\right)$, Ta$_{5}\left(r\right)$, 
and Ta$_{7}\left(r\right)$ respectively. The noise amplification 
is proportional to the number of data utilized in the calculation. This result suggests
a trade-off of raising the accuracy of second order derivative calculation with more data is amplification of the noise variance. 
Thus it is unlikely that the Freeman and Nicholson's and others filters can smooth 
this noise amplification because those filters are 
designed for Ta$_{\text{3}}(r)$ calculation only. One needs new smoothing functions.

\noindent \textbf{The preferred ME filter.}
The new smoothing functions can be obtained through the following procedure. Substituting smoothed data, Eq. (\ref{Phi_s-general}) into second order derivative Ta$_{2M+1}\left(r\right)$ first 
gives a generic expression  
\begin{equation}
D_{s2}\left( r\right) =\sum_{n=-\left( M+z\right) }^{M+z}a_{n}^{\prime }\Phi
\left( r+nh\right) \text{ ,}  \label{D2-g-smooth}
\end{equation}%
where $a_{n}^{\prime }$ is a function of $a^{T}_{m}$ and $p_{l}$. 
The preferred set of $p_{l}$ is then determined by minimizing variance 
$C_{D_{s2}}=\left\langle \left( D_{s2}\left( r\right) -\left\langle D_{s2}\left(
r\right) \right\rangle \right) ^{2}\right\rangle=\sigma ^{2}\sum_{n=-\left( M+z\right) }^{M+z}a_{n}^{\prime 2}%
$.  
It should have high accuracy of 
second order derivative calculation and minimum noise variance simultaneously.
Table \ref{Preferred ME filter} gives the results. 
We define operation based on the ME that utilizes n data points for noise smoothing,S$_{\text{n}}$, and m data for calculating second order derivative, Ta$_{\text{m}}$, as MES$_{\text{n}}$Ta$_{\text{m}}$ in the first column of Table. \ref{Preferred ME filter}. 
Second column records the minimum $C_{D_{s2}}$ value. 
Next five columns list the corresponding $\alpha $ and
the $p_{l}$ value consecutively. The table first shows that when there is
more data, S$_{\text{3}}$, S$_{\text{5}}$, and S$_{\text{7}}$, to be considered for 
smoothing given Ta$_{\text{m}}$, the noise variance $%
C_{D_{s2}}$ will be decreased dramatically. However, when one smoothes data with 
S$_{\text{n}}$ and calculate second order 
derivative using more number of smoothed data points, Ta$_{\text{3}}$, 
Ta$_{\text{5}}$, and Ta$_{\text{7}}$, the noise variance 
$\ C_{D_{s2}}$ will then be amplified as expected. 

\begin{table}[tbp]
\caption{The preferred ME filters, denoted by MES$_{\text{n}}$Ta$_{\text{m}}$.}
\label{Preferred ME filter}\centering
\begin{tabular}{|c|c|c|cccc|}
\hline
& $C_{D_{s2}}\left( \times \sigma ^{2}\right) $ & $\alpha $ & $p_{0}$ & $p_{\pm 1}$ & $p_{\pm 2}$ & $%
p_{\pm 3}$ \\ \hline
MES$_{\text{3}}$Ta$_{\text{3}}$ & 0.28 & 0.031 & 0.42 & 0.28 & 0 & 0 \\ \hline
MES$_{\text{3}}$Ta$_{\text{5}}$ & 0.38 & 0.033 & 0.43 & 0.28 & 0 & 0 \\ \hline
MES$_{\text{3}}$Ta$_{\text{7}}$ & 0.42 & 0.035 & 0.44 & 0.27 & 0 & 0 \\ \hline
MES$_{\text{5}}$Ta$_{\text{3}}$ & 0.048 & 0.017 & 0.29 & 0.23 & 0.12 & 0 \\ \hline
MES$_{\text{5}}$Ta$_{\text{5}}$ & 0.061 & 0.018 & 0.29 & 0.23 & 0.11 & 0 \\ \hline
MES$_{\text{5}}$Ta7 & 0.065 & 0.019 & 0.30 & 0.23 & 0.11 & 0 \\ \hline
MES$_{\text{7}}$Ta$_{\text{3}}$ & 0.014 & 0.011 & 0.22 & 0.19 & 0.12 & 0.06 \\ \hline
MES$_{\text{7}}$Ta$_{\text{5}}$ & 0.017 & 0.012 & 0.23 & 0.2 & 0.12 & 0.05 \\ \hline
MES$_{\text{7}}$Ta$_{\text{7}}$ & 0.019 & 0.012 & 0.23 & 0.2 & 0.12 & 0.05 \\ \hline
\end{tabular}%
\end{table}

\section{Performance analyses}
Instead of applying the proposed ME filter directly to study 
real EEG data for examinations, we will only analyze the filters's performance 
through frequency and noise variance analysis in this work. In addition, 
we will compare it to other filters, mean filter (MF), 
Freeman and Nicholson's (FN), Rappelsberger's (R) 
filters from Table \ref{Compare filter}.

\begin{figure}[tbp]
\centering
\includegraphics[width=3.2in]{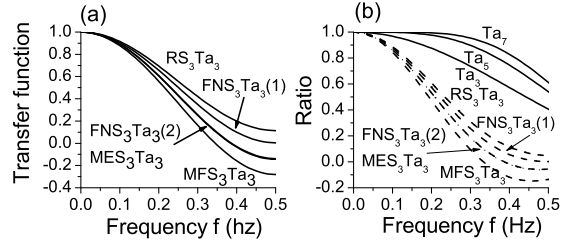} 
\caption{Frequency analysis of the five different 3-point S$_{\text{3}}$ filters in panel (a) 
and different S$_{\text{3}}$,Ta$_{\text{3}}$ in panel (b). }
\label{transfer}
\end{figure}

\noindent \textbf{Frequency analysis.}
The frequency analysis applies the transfer function $H\left( \omega \right)$ for examining
changes of signal's amplitudes after signal being processed by the filters and second 
order derivative calculation \cite{Hamming89}. The
transfer function $H\left( \omega \right)=\exp(-i\omega r)\Phi_{s}(r)=p_{0}+2\sum_{l=1}^{z}p_{l}\cos l\omega $ 
is derived by substituting $\Phi(r)=\exp(i\omega r)$, which is shown to be an eigenfunction 
of a linear time-invariant system, into $\Phi_{s}(r)$ of the Eq.(\ref{D2-g-smooth}), 
where angular frequency $\omega $ equals to $2\pi f$ rotational frequency \cite{Hamming89}. When 
the transfer function is less then one, it indicates signal amplitude being attenuated by 
the filter.
Panel (a) of Fig. \ref{transfer} plots the results of using the different
filters from Table \ref{Compare filter}. 
This figure demonstrates the mean filter has the largest amplitude attenuation and is followed by FNS$_{\text{3}}$(2) and MES$_{\text{3}}$, 
FNS$_{\text{3}}$(1), and RS$_{\text{3}}$ consecutively. 

Similarly, one can compute the transfer function of second order derivative
of smoothed data. Afterward, we compute the ratio of calculated and real 
second order derivative value through
$R\left( \omega \right) =-D_{s2}\left( r\right)/\omega ^{2}=-1/\omega ^{2}\left(%
a_{0}+2\sum_{l=1}^{M+z}a_{l}^{\prime }\cos l\omega \right)$
for explicit comparison of effects of the smoothing. When the ratio $R\left( \omega \right) $ equals to one, the calculated second order derivative of the smoothed data is identical to real value.
Panel (b) of Fig. \ref{transfer} plots not only the calculation through
five filters, 
but also the three second order derivative calculations 
Ta$_{\text{3}}$, Ta$_{\text{5}}$, 
and Ta$_{\text{7}}$ respectively. First of all, frequency attenuation is the smallest in 
Ta$_{\text{7}}$, and is followed by Ta$_{\text{5}}$ and Ta$_{\text{3}}$.
That explains again, why the second order derivative is more accurate when
more data is comprehended for calculation. Given the second order derivative
calculation with fixed number of smoothed data, for example, all S$_{\text{3}}$Ta$_{\text{3}}$ in
this figure, it suggests that the mean filter MFS$_{\text{3}}$Ta$_{\text{3}}$ again has
signal with high frequency attenuated the most among these filters. These
attenuation patterns are similar to the panel (a), which is not surprising because 
smoothing data first and then calculating second order
derivative is identical to calculating second order derivative first and
then smooth the calculation. 



\noindent \textbf{Noise variance analysis.}
It seems the mean filter to be superior to all other choices from the previous analysis. 
However, the following noise variance analysis shows different results. 
We substitute the filters from Table \ref{Compare filter} 
into Eq. (\ref{general-Ta_2M+1}) to compute new coefficient $%
a_{m}^{\prime }$, and evaluate noise variance $C_{D_{s2}}$. The results are listed in Table \ref{filter C and K}. 
Because we design the ME\ filter to mostly reduce noise variance, the analysis shows 
the MES$_{\text{3}}$Ta$_{\text{3}}$ filter to exactly has the best performance among all 
filters. 

Besides, Freeman and Nicholson have found noise in\ field potential
measurements\ to be proportional to a K-value,
$K=\sum_{l=-\left( M+z\right) }^{M+z}\left\vert a_{l}^{\prime }\right\vert 
$ \cite{Freeman75}. 
Thus we calculate the K-value after using different filters and list in the
fourth column of Table \ref{filter C and K}. The K-value shows although 
MES$_{\text{3}}$Ta$_{\text{3}}$ is 
slightly outperformed by the FNS$_{\text{3}}$T$_{\text{3}}$(1), which has the same 
performance as the FNS$_{\text{3}}$T$_{\text{3}}$(2), the ME filter is superior than 
all other filters when more data points are included for smoothing.

\begin{table}[tbp]
\caption{Comparisons of filters. The table lists $a^{\prime}_{n}$ value and 
noise variance $C_{D_{s2}}\left( \times \protect\sigma ^{2}\right) $ in the third column and $K$ value derived by Freeman and Nicholson 
for noise estimate in the fourth column.}
\label{filter C and K}\centering
\begin{tabular}{|l|ccccc|c|c|}
\hline
& $a^{\prime}_{0}$ & $a^{\prime}_{\pm 1}$ & $a^{\prime}_{\pm 2}$ & $%
a^{\prime}_{\pm 3}$ & $a^{\prime}_{\pm 4}$ & $C_{D_{s2}}\left(\times
\sigma^{2}\right)$ & $K$ \\ \hline
MFS$_{\text{3}}$Ta$_{\text{3}}$ & 0 & -0.333 & 0.333 & 0 & 0 & 0.44 & 1.33 \\ \hline
FNS$_{\text{3}}$Ta$_{\text{3}}$(1) & -0.5 & 0 & 0.25 & 0 & 0 & 0.38 & 1.01 \\ \hline
FNS$_{\text{3}}$Ta$_{\text{3}}$(2) & -0.286 & 0.143 & 0.286 & 0 & 0 & 0.28 & 1.13 \\ \hline
RS$_{\text{3}}$Ta$_{\text{3}}$ & -0.623 & 0.082 & 0.229 & 0 & 0 & 0.57 & 1.34 \\ \hline
MES$_{\text{3}}$Ta$_{\text{3}}$ & -0.284 & -0.144 & 0.286 & 0 & 0 & 0.28 & 1.14 \\ \hline\hline
FNS$_{\text{5}}$Ta$_{\text{3}}$ & -0.2 & -0.05 & 0.06 & 0.09 & 0 & 0.073 & 0.63 \\ \hline
FNS$_{\text{7}}$Ta$_{\text{3}}$ & -0.1 & -0.04 & 0.01 & 0.04 & 0.04 & 0.019 & 0.39 \\ \hline
MES$_{\text{5}}$Ta$_{\text{3}}$ & -0.116 & -0.055 & -0.007 & 0.12 & 0 & 0.048 & 0.48 \\ \hline
MES$_{\text{7}}$Ta$_{\text{3}}$ & -0.06 & -0.038 & 0.003 & 0.003 & -0.06 & 0.014 & 0.27 \\ \hline
\end{tabular}%
\end{table}

\noindent \textbf{Discussions.}
The frequency analysis has shown the mean filter MFS$_{\text{3}}$ and the MES$_{\text{3}}$ 
and FNS$_{\text{3}}$(2) mostly attenuate high frequency part in the signal. 
Yet the noise variance analysis, comparison of the $C_{D_{s2}}$ and K value, 
shows the MES$_{\text{3}}$Ta$_{\text{3}}$ and the FNS$_{\text{3}}$Ta$_{\text{3}}$(2) can 
mostly reduce noise variance among all three-point filters. These three analyses suggests
the ME filter not only mostly reduce noise variance in second order derivative 
but also has the ability to mostly attenuate high frequency signal. 

\section{Conclusions}

The filter design delicately depends on one's purpose and properties of a
target system. It sometimes involves some ad hoc empirical guidelines that suit ones need. 
For example, Freeman and Nicholson have designed several filters for their needs \cite{Freeman75}. We propose and
demonstrate an informatic approach to design a low-pass filter based on the method of 
ME. In addition, we examine and compare the preferred ME filter with other well known 
filters in EEG studies through frequency and noise variance analysis. 
The performance of the preferred ME filter MES$_{\text{3}}$Ta$_{\text{3}}$ and 
the FNS$_{\text{3}}$Ta$_{\text{3}}$(2) filter are found to be almost identical. 
Even though the ME\ filter is designed for second order derivative calculation
especially for the EEG\ data in this work only, the approach's systematics and flexibility 
ensures one can design filters that suit other
purposes robustly without introducing ad hoc empirical guidelines.

\section{Acknowledgment}

This work is partially supported by NSC-95-2811-M008-016 to CYT 
and NSC-95-2112-M-008-004 to HCL from the National Science Council, Taiwan ROC.

\end{document}